\documentclass[a4paper]{jpconf}
\usepackage{graphicx,amssymb,amsmath}
\usepackage{graphicx}
\usepackage{bm}

\newcommand{\nt}{$\nu_T=1$}

\newcommand{\Sxx}{$\sigma_{xx}^{||}$}

\begin{document}

\title{Exciton Transport in a Bilayer Quantum Hall Superfluid}

\author{J.P. Eisenstein$^1$,  A.D.K. Finck$^1$, D. Nandi$^1$, L.N. Pfeiffer$^2$, and K.W. West$^2$}

\address{$^1$Department of Physics, California Institute of Technology, Pasadena, CA 91125}
\address{$^2$Department of Electrical Engineering, Princeton University, Princeton, NJ 08544}

\date{\today}

\begin{abstract} 
Bilayer quantum Hall systems at \nt\ support an excitonic ground state.  In addition to the usual charged quasiparticles, this system possesses a condensate degree of freedom: exciton transport.  Detection of this neutral transport mode is facilitated by the use of the Corbino multiply-connected geometry in which charge transport is suppressed.  We here summarize our recent experiments on Corbino devices which directly demonstrate exciton transport across the bulk of the incompressible \nt\ quantum Hall state.
\end{abstract}


Quantum Hall systems are topological insulators.  The bulk of the 2D electron system is an electrical insulator with an energy gap due to either single particle or many-body effects.  At the boundary of the  2D system there are chiral conducting edge states.  The number of such edge states is a topological invariant which determines the Hall resistance of the system. The quantized Hall state which appears in closely-spaced bilayer 2D systems at total Landau level filling \nt\ is no exception to this rule.  In this case Coulomb interactions between the layers are comparable to interactions within each layer.  As a result, at \nt, there is spontaneous interlayer phase coherence; electrons are coherently spread between both layers even in the limit of zero single particle tunneling.  A single edge state circulates at the boundary and the Hall resistance is quantized at $\rho_{xy} = h/e^2$.  Charge transport via this edge state consists of parallel electrical currents in the individual layers.

In addition to ordinary parallel charge transport, the \nt\ bilayer quantum Hall system is unique in possessing an additional, entirely different mode of transport.   This new mode occurs within the system condensate rather than its charged excitations above the energy gap.  The condensate of the \nt\ quantum Hall system may be viewed as a coherent ensemble of interlayer excitons \cite{jpemacd}.  The excitons are analogous to the Cooper pairs in a superconductor, with the important distinction that they are charge neutral.  Being neutral the excitons are not confined to the sample boundaries and should be able to move throughout the bulk of the bilayer system.  Exciton transport is expected to be dissipationless \cite{macdgirvin}, although this has not yet been completely confirmed. Unlike transport of the charged excitations of the system, which amounts to parallel current flow in the two layers, exciton transport is equivalent to oppositely directed, or {\it counterflowing} electrical currents in the layers.

Detecting exciton transport at \nt\ is complicated by several factors.  First, there is the obvious problem of how to control and detect the motion of neutral objects using only electrical means.  Fortunately, this difficulty can be overcome if separate electrical contacts to the two layers are available \cite{sepcon}.  With such contacts it is possible to force oppositely directed currents through the two 2D electron layers and thereby potentially gain access to the condensate degree of freedom.  Second, depending on circumstances, condensate transport may have to compete with ordinary charged quasiparticle transport.  Indeed, early counterflow experiments were performed using Hall bar geometries in which all electrical contacts were placed along the single outside edge of the device \cite{kellogg2,tutuc1,wiersma}.  In this case the highly conducting \nt\ edge state is available for charge transport.  Furthermore, with contacts only on a single edge it was clearly not possible to assert that exciton transport was taking place throughout the bulk of the system.

As first recognized by Tiemann {\it et al.} \cite{tiemann1,tiemann2} and by Su and MacDonald \cite{su}, the multiply-connected Corbino device geometry (roughly an annulus) offers a significant advantage over simply-connected Hall bars for the direct detection of bulk exciton transport.  In this case the edge states on the two boundaries of the annulus are separated by the bulk of the 2D system.  At low temperatures (and sufficiently small bias voltage) charge transport between these edges is heavily suppressed by the quantum Hall energy gap.   At \nt\ however, the bulk of the 2D system remains essentially transparent to exciton transport.  We thus arrive at the remarkable conclusion that even though charge transport via parallel currents in the two layers across the bulk is suppressed, oppositely directed layer currents should be able to move freely by making use of the condensate degree of freedom. 

In a recent experiment, Finck {\it et al.} \cite{ finck} presented convincing evidence for exciton transport across the bulk of a Corbino ring at \nt.  With four ohmic contacts on the outside rim of the annulus (two on each 2D layer) and two more on the inner rim (one on each layer) a variety of transport measurements could be performed.  At low temperatures and sufficiently small effective interlayer separation ($d/\ell \sim 1.5$, with $d$ the physical separation of the layers and $\ell$ the magnetic length) the \nt\ bilayer quantum Hall effect was robust in their sample. This was clearly evident from, among other signatures, the exponentially suppressed parallel charge conductivity \Sxx\ across the annulus.

In order to search for bulk exciton or counterflow conductivity a bias voltage $V$ was applied between top and bottom layer contacts (the source and the drain) on the outer rim of the annulus and the resulting current $I_1$ was recorded.  (The two additional contacts on the outer rim were used to provide a 4-terminal measurement of the interlayer voltage difference, $V_i$.) On the inner rim the remaining top and bottom layer contacts could be connected together via a shunt resistor $R_s$ or left disconnected.   With the shunt disconnected this is a simple interlayer tunneling configuration.  With the shunt in place additional current might flow from source to drain if counterflowing layer currents cross the otherwise insulating bulk of the 2D system and pass through the shunt.  The current $I_2$ passing through the shunt is easily determined from the measured voltage drop across $R_s$.  Figure 1a schematically illustrates the configuration with the shunt connected.
\begin{figure}[t]
\begin{center}
\includegraphics[width=5.75in] {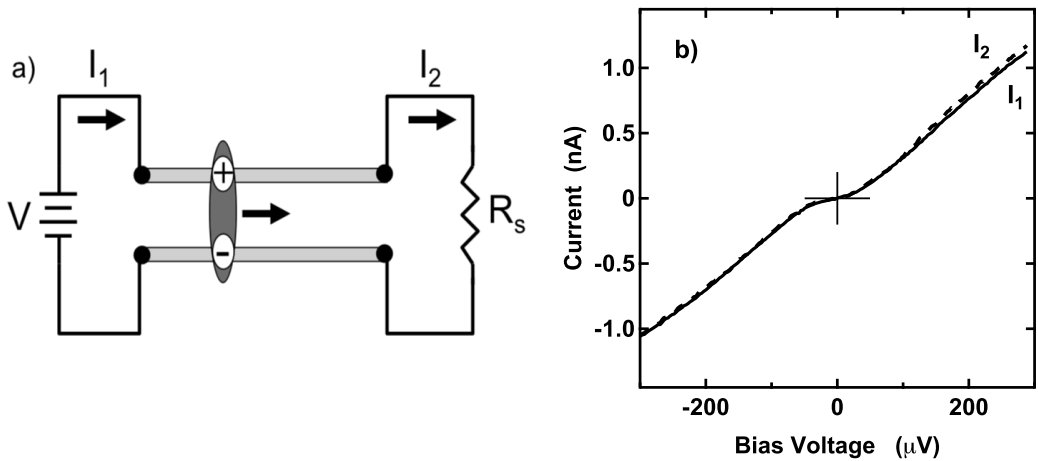}
\end{center}
\caption{a) Schematic illustration of the circuit geometry used by Finck {\it et al.} \cite{finck} to detect exciton transport at \nt.  The bilayer electron system is shown in cross-section, with the solid dots on the left being on the outer rim of the Corbino annulus and the solid dots on the right being on the inner rim.  In the ideal situation shown, only exciton transport is occurring; there is no interlayer tunneling and no charged quasiparticle transport.  In these ideal conditions, the currents $I_1$ and $I_2$ are precisely equal.  Additional resistances (due to contacts, etc.) have been omitted for clarity.  b) Measured bias and shunt currents at \nt, with $d/\ell = 1.5$ and $T \sim 20$ mK and the sample tilted to suppress coherent tunneling. The near perfect equality of $I_1$ and $I_2$ demonstrates the dominance of exciton transport across the insulating bulk of the 2D system.}
\label{fig1}
\end{figure}

With the magnetic field applied perpendicular to the 2D planes and the shunt disconnected, Finck {\it et al.} observed the well-known Josephson-like tunneling response of the bilayer system \cite{spielman1}.  As the bias voltage $V$ was increased from zero the current $I_1$ tunneling between the layers steadily increased up to a critical current of about $I_c \sim 1.5$  nA.  Simultaneous measurements of the interlayer voltage $V_i$ showed that it remained very close to zero for currents below the critical current.  When the bias voltage was increased further, the interlayer voltage abruptly jumped to a finite value and the current dropped below the critical value.  As observed previously \cite{spielman1,tiemann2}, these two states strongly resemble the superconducting and resistive states of a conventional Josephson junction.

Connecting the shunt on the inner rim had no effect on the above results until the system was driven into the resistive state.  For small bias voltages, where $I_1<I_c$ and $V_i \sim 0$, no current was observed to flow through the shunt.  In this situation it would appear that all of the bias current was simply tunneling between the layers and not counterflowing across the bulk to the shunt.  At higher bias, in the resistive state, current does begin to flow through the shunt.  

Expecting these results, Finck {\it et al.} then rotated the sample relative to the magnetic field so that a significant in-plane component $B_{||}\sim 1$ T was present in addition to the perpendicular component $B_{\perp}\sim 2$ T used to establish \nt.  This drastically reduces the critical current for coherent Josephson-like tunneling \cite{spielman2} but otherwise has little effect on the \nt\ QHE.  Indeed, with the shunt disconnected very little current was observed to flow in response to the bias voltage $V_b$.  In contrast, when the interlayer shunt was connected large currents flowed and careful measurements revealed that the shunt current $I_s$ almost perfectly matched the source-drain current $I$.  In this case essentially all of the current is counterflowing across the bulk of the Corbino ring.  This result constitutes direct proof of exciton transport across the insulating bulk of the \nt\ bilayer quantum Hall system.  Further measurements showed that the magnitude of this excitonic current was overwhelmingly dominated by the various extrinsic resistances in the circuit; any dissipation in the exciton transport itself was undetectable.

In the above measurements the counterflowing currents are taking place at finite interlayer voltage, $V_i$.  This voltage is inevitable in the circuit arrangement Finck {\it et al.} employed, if only because of the non-zero resistance of the shunt connection.  Hence, the exciton transport was accompanied by a time-dependent condensate phase.  We stress that this does not imply that dissipation is occurring in the exciton channel. So long as the interlayer voltage is uniform across the device (and the measurements suggested that this was the case), the order parameter winds at a constant rate with no phase slips being generated.

An alternative method for inducing exciton transport across the Corbino ring is to use a Coulomb drag configuration \cite {kogan,su}.  In this case the source and drain contacts for the bias circuit are on the same 2D layer (the drive layer) but on opposite rims of the annulus.  Meanwhile, inner and outer rim contacts on the opposite layer (the drag layer) are connected together through an external resistor, $R_D$.  This arrangement is schematically illustrated in Fig. 2a.  In the absence of strong interlayer correlations, driving a current $I_1$ through the drive layer will induce only a very tiny current $I_2$ in the drag layer.  In contrast, at \nt\ interlayer correlations are very strong and exciton transport within the condensate offers an alternative: the current $I_2$ in the drag layer will precisely match the current $I_1$ in the drive layer, but be oppositely directed, as excitons flow uniformly across the Corbino ring.  Furthermore, in the absence of dissipation in the exciton channel, the magnitude of these currents will determined only by the bias voltage $V_b$ and the sum $R_{tot}$ of all extrinsic resistances in the drive and drag loops. (We include in $R_{tot}$ quantum Hall contact resistances of order $h/e^2$; see Su and MacDonald \cite{su}.)  Unlike the experiment of Finck {\it et al.}, exciton transport in this case is generated without any explicit electrical connection between the two layers.

\begin{figure}[b]
\begin{center}
\includegraphics[width=5.75in] {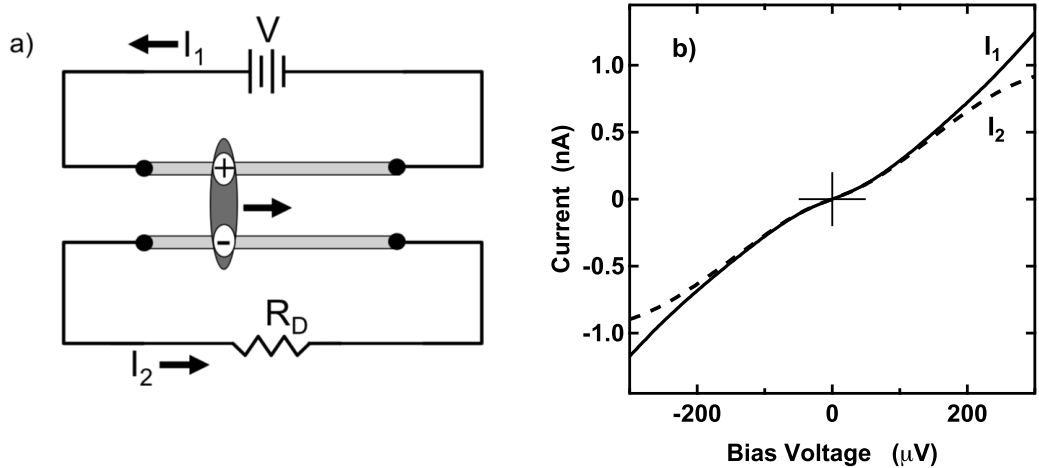}
\end{center}
\caption{a) Schematic illustration of the circuit geometry for Corbino Coulomb drag measurements used by Nandi {\it et al.} \cite{nandi}. b) Drive and drag currents, $I_1$ and $I_2$, at \nt\ and $d/\ell = 1.5$ and $T \sim 20$ mK.  Under these conditions the excitonic phase is robust.  At small bias voltages equal, but oppositely directed, currents flow in the two layers:  $I_1 \approx I_2$.  At higher bias voltages the two currents separate.  This is due to enhanced charge transport across the bulk of the Corbino annulus. The sample has again been tilted relative to the magnetic field in order to suppress interlayer tunneling in the excitonic phase.}
\label{fig2}
\end{figure}

Recent experiments by Nandi {\it et al.} \cite{nandi} have verified this scenario of exciton-mediated Coulomb drag.  Figure 2b illustrates their basic findings, with the drive and drag currents, $I_1$ and $I_2$, plotted versus bias voltage $V$.  These data were obtained at \nt\ and $T \sim 20$ mK and $d/\ell = 1.5$.   At this effective layer separation the excitonic \nt\ QHE state is well-developed and measurements of the parallel charge transport conductivity, \Sxx, show it to be heavily suppressed.  As before, the sample has been tilted relative to the applied magnetic field in order to suppress coherent interlayer tunneling.  As the figure shows, for bias voltages below about 150 $\mu$V, equal currents flow across the bulk of the Corbino ring in the drive and drag loops; i.e. $I_1 \approx I_2$.  In this situation, Coulomb drag is ``perfect'' and reflects a uniform flow of excitons across the bulk of the 2D sample.  

We emphasize that when the drag layer is not closed upon itself via the resistor $R_D$ virtually no current flows in $either$ layer in response to small bias voltages.  Indeed, were significant current to flow in the drive layer without a compensating opposite current in the drag layer, it would conflict with the observation that \Sxx\ is extremely small.  

Above about $V \approx 150$ $\mu$V the drive current begins to exceed the drag current.  In this regime net charge is being transported across the Corbino ring.  Detailed measurements \cite{nandi} show that \Sxx\ grows significantly as the \nt\ QHE begins to ``breakdown'' with increasing bias voltage $V$.  In fact, we find it possible to convincingly simulate the behavior of $I_1$ and $I_2$ with increasing $V$ via independent measurements of \Sxx\ in the non-linear regime.  A similar analysis also explains the deviations from perfect Coulomb drag which arise from the temperature dependence of \Sxx.

Finally, we note that as expected, when the effective layer separation is increased to $d/\ell = 2.3$ and interlayer coherence is lost, essentially no current flows in the drag layer even though current is readily transported across the drive layer.  In this case the bilayer system consists of essentially two independent 2D layers each at $\nu = 1/2$.

In summary, exciton transport across the insulating bulk of the \nt\ bilayer QHE state has been observed in two related, but distinct, transport experiments in Corbino geometry.  While these experiments are so far consistent with dissipationless transport of the exciton ensemble, they do not place stringent upper bounds on it.  Future multi-terminal measurements will address this important question. 

This work was supported by the NSF under grant DMR-1003080.
\newline
\newline

\end{document}